\documentclass[english,nofootinbib,aps]{article}
\usepackage[T1]{fontenc}
\usepackage[latin9]{inputenc}
\usepackage[a4paper]{geometry}
\usepackage{color}
\usepackage{babel}
\usepackage{mathrsfs}
\usepackage{amsmath}
\usepackage{amssymb}
\usepackage{graphicx}
\usepackage[unicode=true,pdfusetitle,
 bookmarks=true,bookmarksnumbered=false,bookmarksopen=false,
 breaklinks=false,pdfborder={0 0 1},backref=false,colorlinks=false]
 {hyperref}

\makeatletter

\newcommand{\lyxdot}{.}

\newcommand{\lyxaddress}[1]{
	\par {\raggedright #1
	\vspace{1.4em}
	\noindent\par}
}

\@ifundefined{date}{}{\date{}}

\usepackage{babel}

\@ifundefined{showcaptionsetup}{}{%
 \PassOptionsToPackage{caption=false}{subfig}}
\usepackage{subfig}
\makeatother

\begin{document}
\title{Higher-order nonclassicalities in hybrid coherent states}
\author{{\normalsize{}Sandip Kumar Giri$^{1}$, Biswajit Sen$^{2,}$}\thanks{bsen75@yahoo.co.in}}
\maketitle

\lyxaddress{\begin{center}
$^{1}$Department of Physics, Panskura Banamali College, Panskura-721152,
India\\
 $^{2}$Department of Physics, Vidyasagar Teachers' Training College,
Midnapore-721101, India
\par\end{center}}
\begin{abstract}
{\normalsize{}In the development of quantum technologies, nonclassical
states have been playing a pivotal role, as quantum advantage cannot
be obtained without appropriate utilization of nonclassicality. In
the present work, we consider a hybrid coherent state (HCS), which
is a coherent superposition of the single-photon-added coherent (SPAC)
state and a coherent state (CS). Here, we report higher-order nonclassical
properties of HCS with a specific focus on higher-order squeezing
and higher-order antibunching. It's shown that HCS is experimentally
realizable, and this engineered quantum state can be used to produce
quantum states with desired higher-order nonclassical properties.}{\normalsize\par}
\end{abstract}

\section{{\normalsize{}Introduction}}

The field of quantum state engineering has been considerably advanced
in the recent past, and several engineered quantum states have been
produced \cite{nonclassical SPAC 2,superposition-coherentstate}.
Further, applications of the nonclassical engineered states have been
reported in quantum teleportation, quantum communication, and quantum
information theory \cite{engineered state application 2}. Over the
last few decades, various schemes have been proposed to develop different
kinds of quantum states, including squeezed state, Fock state, etc.
Furthermore, some new kinds of states that are generated by adding
or subtracting a photon with a coherent state have been found to show
very interesting properties. For example, we may mention that the
photon-added coherent state was first introduced by Agarwal and Tara
\cite{PAC state} in 1991. It is a state intermediate between the
Fock state and the coherent state, and it shows various nonclassical
features \cite{PAC state,nonclassical-spac} like photon antibunching,
squeezing, etc. Now, the single-photon-added coherent state, indicated
by the symbol $\left|\alpha,1\right\rangle $ is produced when a photon
is added to a coherent state. Such a state was experimentally produced
by Zavatta et al. \cite{tomography-spac-firstexp} using basic optical
processes such as beam splitting, frequency down-conversion \cite{sivakumar-spac-parametric},
and homodyne detection. Due to the distinctive characteristics of
the SPAC state, it has a variety of uses in quantum optics and quantum
information technology, such as quantum key distribution \cite{qkd-spac-1,qkd-spac},
quantum communication \cite{quantum communication-spac}, quantum
sensing \cite{spac-quantumsesing}, and quantum metrology \cite{spac-metrology}.
The SPAC state serves as a versatile test-bed for studying the strange
nature of quantum mechanics. It shows the nonclassical features, such
as quadrature squeezing \cite{squeezing-spac}, negativity of the
Wigner function \cite{spac-wigner-negativity}, sub-Poissonian photon
statistics \cite{PAC state}, etc. Recently, Yususf Tarek et al. \cite{Turek1}
proposed a new kind of state termed HCS, which is generated by adding
a coherent state to a single-photon-added coherent state. The HCS
has several advantages over the SPAC state \cite{Turek1}. The nonclassical
features of HCS, such as photon statistics, quadrature squeezing,
and amplitude squared squeezing, have been extensively studied \cite{Turek1},
although the higher-order nonclassicalities of HCS have not been studied
yet. It's known that the higher-order nonclassicalities of engineered
coherent states often exhibit more pronounced nonclassical effects
than their lower-order counterparts, and they can be used for various
quantum information processing tasks, such as quantum computation,
quantum communication, quantum cryptography, etc. \cite{higher-order-spac}.
For example, higher-order squeezing is shown to enhance different
aspects of quantum computing technology, such as quantum communication
and imaging \cite{quantumimaging-squeezing,communication-highsqueez,laiho-higherorsqeezing}.
Keeping these in mind, we aim to study the higher-order nonclassical
effects (e.g., higher-order Hong-Mandel type squeezing, higher-order
antibunching) in HCS.

The rest of the paper is organized as follows: As this paper aims
to study some nonlclassical properties of HCS, Section \ref{sec:Hybrid-coherent-state}
is dedicated to the description of HCS. Subsequently, in Section \ref{sec:Possible-experimental-generation},
a method for experimental generation of HCS is described with an appropriate
schematic diagram of the possible setup to establish that the results
of the present work are experimentally realizable with the existing
technology. In Section \ref{sec:Higher-order-Squeezing}, it's shown
that HCS exhibits higher-order squeezing, and the amount of maximum
squeezing increases with the order of squeezing. A similar nature
is observed for higher-order antibunching and the same is described
in Section \ref{sec:Higher-order-antibunching}. Finally, the paper
is concluded in Section \ref{sec:Conclusions}.

\section{Hybrid coherent state\label{sec:Hybrid-coherent-state}}

An interesting engineered quantum state is HCS \cite{Turek1}, which
is a superposition between a closest to classical state, called a
coherent state, and a nonclassical (SPAC) state where the amplitudes
of the CS and the CS on which a photon is added (equivalently, the
creation operator $a^{\dagger}$ is applied) are the same. Such a
superposition can be written as 
\begin{equation}
\begin{array}{lcl}
\left|\varPsi\right\rangle  & = & c_{1}\left|\alpha\right\rangle +c_{2}e^{i\varphi}\left|\alpha,1\right\rangle ,\end{array}\label{superpositionstate}
\end{equation}
where $\left|\alpha\right\rangle $ and $\left|\alpha,1\right\rangle $
represent CS and SPAC state, respectively, and $c_{i\in\left\{ 1,2\right\} }$are
their amplitudes, respectively. The CS and the SPAC states are also
separated by a phase angle $\varphi$. For $c_{1}=0$ and $c_{2}=0$
the state $\left|\varPsi\right\rangle $ reduces to SPAC state and
CS, respectively. Thus, it can be treated as an intermediate state.
The SPAC state is a well-studied nonclassical state, and it's easy
to visualize its nonclassical nature as the addition of a photon leads
to the creation of a\textcolor{red}{{} }\textcolor{green}{hole} in the
photon number distribution, which is a witness of nonclassicality
\cite{anirban-review,hole-burning}. The question is, whether the
nonclassicality present in the SPAC state remains preserved in its
superposition with CS? We will address this question in this paper.
However, before we do so, we would like to note that the combined
properties of CS and the SPAC state in HCS would depend on the values
of the superposition parameters $c_{1}$ and $c_{2}$. Also, the SPAC
state reduces to a single-photon state for $\left|\alpha\right|\rightarrow0$
and a coherent state for $\left|\alpha\right|\rightarrow\infty$.
The coherent state $\left|\alpha\right\rangle $ is the eigenstate
of the annihilation operator $a$ with eigen value $\alpha$, i.e.,
$a\left|\alpha\right\rangle =\alpha\left|\alpha\right\rangle .$ The
complex eigen value $\alpha=\left|\alpha\right|e^{i\zeta}$, where
$\zeta$ is a phase angle and a real number. The coherent state in
the Fock basis $\left|n\right\rangle $ can be represented as 
\begin{equation}
\begin{array}{lcl}
\left|\alpha\right\rangle  & = & e^{-\frac{\left|\alpha\right|^{2}}{2}}\sum_{n=0}^{\infty}\frac{\alpha^{n}}{\sqrt{n!}}\left|n\right\rangle \\
 & = & e^{\left(\alpha a^{\dagger}-\alpha^{*}a\right)}\left|0\right\rangle \\
 & = & \mathcal{D\left(\alpha\right)}\left|0\right\rangle ,
\end{array}\label{eq:coherentstate}
\end{equation}
 where $\mathcal{D\left(\alpha\right)}$ is the displacement operator. 

\section{Possible experimental generation of a hybrid coherent state\label{sec:Possible-experimental-generation}}

\begin{figure}
\begin{centering}
\includegraphics{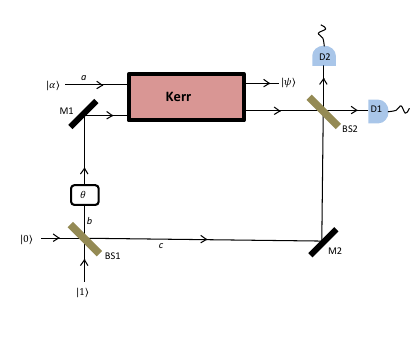}\caption{The scheme for the generation of HCS\label{fig:Scheme-for-generation}}
\par\end{centering}
\end{figure}

The HCS can be prepared through the coupling of a single-photon Mach-Zehnder
interferometer with a nonlinear optical Kerr medium \cite{Turek1,Matsuoka}.
A single-photon state $\left|1\right\rangle $ and a vacuum state
$\left|0\right\rangle $ are injected into the two input ports of
the BS1, which is a 50:50 beam splitter as shown in Fig. \ref{fig:Scheme-for-generation}.
The resulting state at the output ports $b$ and $c$ of BS1 will
be 
\begin{equation}
\left|Out\right\rangle _{BS1}=\frac{1}{\sqrt{2}}\left(\left|10\right\rangle _{bc}+i\left|01\right\rangle _{bc}\right),\label{eq:BS1}
\end{equation}

where we have used the obvious notation $\left|1\right\rangle _{b}\left|0\right\rangle _{c}\equiv\left|10\right\rangle _{bc}$.
The state in arm $b$ picks up a phase shift $\theta$. So, after
phase shifting, we have the state
\begin{equation}
\left|Out\right\rangle _{BS1,\theta}=\frac{1}{\sqrt{2}}\left(e^{i\theta}\left|10\right\rangle _{bc}+i\left|01\right\rangle _{bc}\right).\label{eq:BS1, Theta}
\end{equation}
 A coherent state $\left|\alpha\right\rangle $ is present in arm
$a$. So, before the cross-Kerr interaction, we have the input state
as
\begin{equation}
\left|In\right\rangle _{ck}=\left|Out\right\rangle _{BS1,\theta}\left|\alpha\right\rangle _{a}.\label{eq:Cross-Kerr input}
\end{equation}
Now, the state in arm $b$ is coupled with a coherent beam $\left|\alpha\right\rangle $
in arm $a$ through a nonlinear cross-Kerr medium, which induces a
cross-Kerr phase modulation (XPM). The related interaction Hamiltonian
is given by $H_{ck}=\hbar Ka^{\dagger}ab^{\dagger}b$, where $K$
is proportional to $\chi^{(3)}$, the third-order nonlinearity of
the Kerr medium. The action of the Kerr medium yields the output state
as 
\begin{equation}
\left|Out\right\rangle _{ck}=e^{-\frac{iH_{ck}t}{\hbar}}\left|In\right\rangle _{ck}=e^{-i\varPhi a^{\dagger}ab^{\dagger}b}\left|In\right\rangle _{ck},\label{eq:}
\end{equation}
where $\varPhi=Kt$ is the phase shift of XPM in the interaction time
$t$. For weak XPM, the phase shift angle $\varPhi<<1$. Under such
condition, we can approximate $e^{-i\varPhi a^{\dagger}ab^{\dagger}b}\approx\left(1-i\Phi a^{\dagger}ab^{\dagger}b\right)$.
So, under the weak XPM, we have the output state as 
\begin{equation}
\left|Out\right\rangle _{ck}=\frac{1}{\sqrt{2}}\left[\left(1-i\Phi\alpha a^{\dagger}\right)e^{i\theta}\left|10\right\rangle _{bc}\left|\alpha\right\rangle _{a}+i\left|01\right\rangle _{bc}\left|\alpha\right\rangle _{a}\right].\label{eq:Output Cross Kerr}
\end{equation}
Now, the BS2 is a variable beam splitter, which causes the transformations\label{eq:Cross-Kerr output 1}
\begin{equation}
\begin{array}{rcl}
\left|10\right\rangle _{bc} & \stackrel{BS2}{\rightarrow} & \acute{t}\left|10\right\rangle _{D1,D2}+i\acute{r}\left|01\right\rangle _{D1,D2},\\
\left|01\right\rangle _{bc} & \stackrel{BS2}{\rightarrow} & i\acute{r}\left|10\right\rangle _{D1,D2}+\acute{t}\left|01\right\rangle _{D1,D2},
\end{array}\label{eq:BS2 transformation}
\end{equation}
where the real numbers $\acute{t}$ and $\acute{r}$ are the transmissivity
and reflectivity of the BS2, respectively, with $\acute{t}^{2}+\acute{r}^{2}=1$.
The output ports of the BS2 are $D1$ and $D2$. Putting Eq. (\ref{eq:BS2 transformation})
in Eq. (\ref{eq:Output Cross Kerr}), we get the output state of BS2
as 
\begin{equation}
\left|Out\right\rangle _{BS2}=\frac{1}{\sqrt{2}}\left[\left(\acute{t}e^{i\theta}-i\acute{t}e^{i\theta}\varPhi\alpha a^{\dagger}-\acute{r}\right)\left|\alpha\right\rangle _{a}\left|10\right\rangle _{D1,D2}+\left(i\acute{r}e^{i\theta}+i\acute{t}+\acute{r}e^{i\theta}\varPhi\alpha a^{\dagger}\right)\left|\alpha\right\rangle _{a}\left|01\right\rangle _{D1,D2}\right].\label{eq:Output BS2}
\end{equation}
When the single photon is detected at $D1$, the coherent state $\left|\alpha\right\rangle $
in mode $a$ is projected into the state
\begin{equation}
\left|\varPsi\right\rangle =\frac{1}{\sqrt{2}}\left[\left(\acute{t}e^{i\theta}-\acute{r}\right)\left|\alpha\right\rangle _{a}-i\acute{t}e^{i\theta}\varPhi\alpha a^{\dagger}\left|\alpha\right\rangle _{a}\right].\label{eq:Final state}
\end{equation}
We get the HCS, which is the linear superposition of the coherent
state $\left|\alpha\right\rangle $ and the single-photon-added coherent
state $a^{\dagger}\left|\alpha\right\rangle $ when $D1$ fires. The
amplitude and phase of the superposition of the CS and the SPAC state
can be manipulated by adjusting $\acute{r}$ or $\acute{t}$, $\varPhi$,
$\alpha$ and $\theta$. In its general form, we can write the HCS
as

\begin{equation}
\left|\varPsi\right\rangle =\mathscr{N}\left[\sqrt{\epsilon}\left|\alpha\right\rangle +\sqrt{1-\epsilon}e^{i\varphi}a^{\dagger}\left|\alpha\right\rangle \right],\label{eq:HCS}
\end{equation}
where $\epsilon$ determines the amplitudes of the CS and SPAC states
in HCS and $\varphi$ is the phase angle between the CS and the SPAC
states. The normalization constant $\mathcal{N}$ can be expressed
as 
\begin{equation}
\mathscr{N}=\left[1+2\sqrt{\epsilon(1-\epsilon)}\,Re\left[\alpha e^{-i\varphi}\right]+\left(1-\epsilon\right)\left|\alpha\right|^{2}\right]^{-\frac{1}{2}}.\label{eq:Normalization const.}
\end{equation}
The parameter $\epsilon$ is real and has its value $0\leq\epsilon\leq1$
determines the relative contribution of CS and the SPAC state in the
HCS. For $\epsilon=0$ and $\epsilon=1$, the HCS transforms to the
SPAC state and the CS, respectively. In what follows, the expression
for the HCS in Eq. (\ref{eq:HCS}) is exploited to investigate higher-order
squeezing and higher-order antibunching.

\section{{\normalsize{}Higher-order Squeezing\label{sec:Higher-order-Squeezing}}}

A nonclassical phenomenon known as squeezing involves reducing the
uncertainty in a particular quadrature at the cost of increasing the
uncertainty in a noncommutating quadrature. Higher-order squeezing
is a notion in quantum optics that extends the idea of squeezing past
the uncertainty in the quadrature measured via variance. Two separate
approaches are typically used to study higher-order squeezing \cite{Hong1,hong2,hillery}.
Hong and Mandel originally proposed an approach for defining higher-order
squeezing in 1985 \cite{Hong1,hong2}, which is a natural generalization
of second-order squeezing. In contrast, the second notion of higher-order
squeezing was introduced in 1987 by Hillary \cite{hillery}, where
the reduction of the variance of the amplitude-powered quadrature
variable with respect to its coherent state counterpart was considered
as manifestation of higher-order squeezing. In this section, we will
investigate the higher-order Hong-Mandel-type squeezing of the hybrid
coherent state. In order to investigate the higher-order squeezing,
we consider two quadrature components of the field as $X_{\psi}$
and $X_{\psi+\frac{\pi}{2}}$. Let the commutation relation between
the quadratures is $\left[X_{\psi},X_{\psi+\frac{\pi}{2}}\right]=iC$,
where $C$ is a constant. In the course of the work, we will assume
that $\hbar=1.$ Now, the general expression for the $2n$-order variance
of any quadrature $X_{\psi}$ can be written as \cite{Hong1}

\begin{equation}
\begin{array}{lcl}
\left\langle \left(\Delta X_{\psi}\right)^{2n}\right\rangle  & = & \left\langle \colon\left(\Delta X_{\psi}\right)^{2n}\colon\right\rangle +\frac{2!}{1!}\binom{2n}{2}\frac{C}{4}\left\langle \colon\left(\Delta X_{\psi}\right)^{2n-2}\colon\right\rangle \\
 & + & \frac{4!}{2!}\binom{2n}{4}\left(\frac{C}{4}\right)2\left\langle \colon\left(\Delta X_{\psi}\right)^{2n-4}\colon\right\rangle +\cdots+\left(2n-1\right)!!\left(\frac{C}{2}\right)^{n}\\
 & = & S_{\psi}^{(2n)}+\left(2n-1\right)!!\left(\frac{C}{2}\right)^{n},
\end{array}\label{eq:Squeezing 2n_1}
\end{equation}
where the notation $::$ represents the normal ordering where the
creation operators are consistently maintained to the left of the
annihilation operators and

\begin{equation}
S_{\psi}^{(2n)}=\sum_{m=0}^{n-1}\frac{(2n)!}{(2m+2)!(n-m-1)!}\left(\frac{C}{4}\right)^{n-m-1}\left\langle \colon\left(\Delta X_{\psi}\right)^{2m+2}\colon\right\rangle .\label{eq:squeezing 2n_2}
\end{equation}
In the Eqs. (\ref{eq:Squeezing 2n_1}) and (\ref{eq:squeezing 2n_2}),
$m$ and $n$ are the positive integers. Now, the falls of the $2n$-order
fluctuation of a quadrature are below its CS value, indicating the
$2n$-order squeezing of that quadrature. So, the essential requirement
for $2n$-order Hong-Mandel type squeezing in $X_{\psi}$ quadrature
is \cite{hong2,verma}
\begin{equation}
\left\langle \left(\Delta X_{\psi}\right)^{2n}\right\rangle <\left(2n-1\right)!!\left(\frac{C}{2}\right)^{n}.\label{eq:Squeezing 2n_3}
\end{equation}
From the Eqs. (\ref{eq:Squeezing 2n_1}), (\ref{eq:squeezing 2n_2}),
and (\ref{eq:Squeezing 2n_3}) it is clear that the signature of $2n$-order
squeezing of the quadrature $X_{\psi}$ will be seen if $S_{\psi}^{(2n)}<0$.
Now, for the present work, we define the quadrature operators as $X_{\psi}=\frac{1}{\sqrt{2}}\left(a^{\dagger}e^{i\psi}+ae^{-i\psi}\right)$
and $X_{\psi+\frac{\pi}{2}}=\frac{i}{\sqrt{2}}\left(a^{\dagger}e^{i\psi}-ae^{-i\psi}\right)$.
Using the commutation relation defined above, we obtain the most general
expression of $S_{\psi}^{(2n)}$ as
\begin{equation}
\begin{array}{lcl}
S_{\psi}^{(2n)} & = & \sum_{m=0}^{n-1}\frac{(2n)!}{(n-m-1)!(2m+2)!}\left(\frac{C}{4}\right)^{(n-m-1)}\sum_{p=0}^{2m+2}\binom{2m+2}{p}\left(\frac{1}{\sqrt{2}}\right)^{2m+2}\\
 & \times & \sum_{l=0}^{2m+2-p}\binom{2m+2-p}{l}\left\langle a^{\dagger(2m-p-l+2)}a^{l}\right\rangle e^{i(2m-p-2l+2)\psi}\left(-1\right)^{p}\\
 & \times & \sum_{j=0}^{p}\binom{p}{j}\left\langle a^{\dagger}\right\rangle ^{p-j}\left\langle a\right\rangle ^{j}e^{i(p-2j)\psi}.
\end{array}\label{eq:Squeezing 2n_4}
\end{equation}
In our present study, $C=1.$ The expectation values of the $\left\langle a^{\dagger n}a^{m}\right\rangle $
and $\left\langle a\right\rangle $ for our HCS as defined in Eq.
(\ref{eq:HCS}) are obtained in order to investigate the possibilities
of observing the higher-order squeezing. We derive

\begin{equation}
\begin{array}{lcl}
\left\langle a^{\dagger n}a^{m}\right\rangle  & = & \mathscr{N}^{2}\alpha^{*n-1}\alpha^{m-1}\left[\epsilon\left|\alpha\right|^{2}+\sqrt{\epsilon(1-\epsilon)}\left\{ e^{i\varphi}\left(m\alpha^{*}+\left|\alpha\right|^{2}\alpha^{*}\right)\right.\right.\\
 &  & \left.\left.+e^{-i\varphi}\left(n\alpha+\left|\alpha\right|^{2}\alpha\right)\right\} +\left(1-\epsilon\right)\left\{ nm+(n+m+1)\left|\alpha\right|^{2}+\left|\alpha\right|^{4}\right\} \right],
\end{array}\label{eq:Average(a dagger n a m)}
\end{equation}

\begin{equation}
\left\langle a\right\rangle =\mathscr{N}^{2}\left[\epsilon\alpha+\sqrt{\epsilon(1-\epsilon)}\left\{ e^{i\varphi}\left(1+\left|\alpha\right|^{2}\right)+e^{-i\varphi}\alpha^{2}\right\} +\left(1-\epsilon\right)\alpha\left(2+\left|\alpha\right|^{2}\right)\right].\label{eq:average (a)}
\end{equation}
Using Eqs. (\ref{eq:Average(a dagger n a m)}) and (\ref{eq:average (a)}),
we evaluate Eq. (\ref{eq:Squeezing 2n_4}) to study the $2n$-order
Hong-Mandel type squeezing for HCS. In order to get the flavour of
the nonclassicality manifested through Eq. (\ref{eq:Squeezing 2n_4}),
we have plotted 
\begin{figure}
\begin{centering}
\subfloat{\centering{}\includegraphics{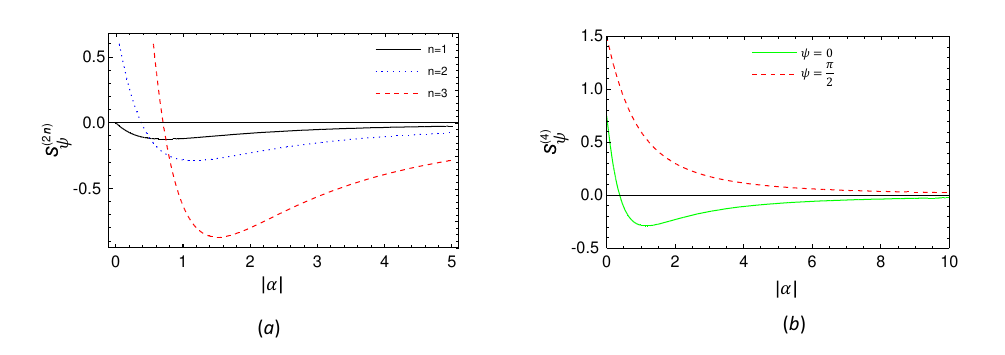}}
\par\end{centering}
\centering{}\caption{(Colour online) Plot of higher-order squeezing with $\left|\alpha\right|$:
(\textit{a}) in $X_{\psi}$quadrature for $n=1$, $2$, and $3$ with
$\epsilon=0.5$; (\textit{b}) in both quadratures for $\epsilon=0.5$.}
\label{Plot of squeezing}
\end{figure}
 the variation of $S_{\psi}^{(2n)}$ with $\left|\alpha\right|$ in
Fig. \ref{Plot of squeezing}. The signature of higher-order squeezing
with different orders is evident in Fig. \ref{Plot of squeezing}(\textit{a}),
which also demonstrates that the degree of squeezing rises with order.
Fig. \ref{Plot of squeezing}(\textit{b}) shows that the squeezing
in one quadrature prohibits the squeezing in another quadrature as
expected due to Heisenberg's uncertainty relation. Fig. \ref{Plot of squeezing}
clearly shows that when $\left|\alpha\right|$ rises, the plot approaches
a coherent state. The solid line, dashed line, dotted line, and dot-dashed
line in Fig. \ref{fig3:4thorder squeezing} 
\begin{figure}
\centering{}\includegraphics{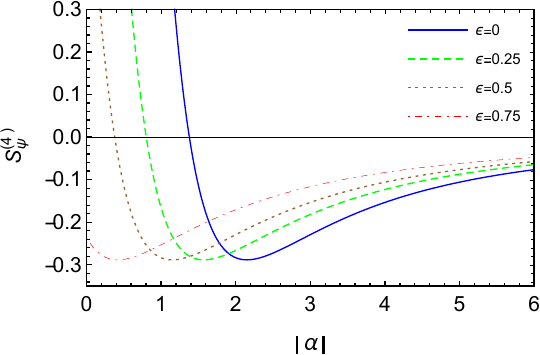} \caption{(colour online) Plot of $4^{th}$ order squeezing with $\left|\alpha\right|$
for $\epsilon=0,\,0.25,\,0.5,$ and $0.75$.}
\label{fig3:4thorder squeezing}
\end{figure}
 represent the $4^{th}$-order squeezing for $\epsilon=0,$ $0.25,$
$0.5$, and $0.75,$ respectively. Hence, the value of the mixing
parameter $\epsilon$ determines the nature of the quadrature squeezing.
As the $\left|\alpha\right|$ increases, the HCS goes closer to the
coherent state. The HCS loses the nonclassical squeezing property
for $\epsilon=1.$ Interestingly, for single photon ( $\left|\alpha\right|\rightarrow0$)
or few photons ( $\left|\alpha\right|\rightarrow$small), the HCS
($\epsilon\neq0$) is observed to show higher-order squeezing, but
is not observed if we take only the SPAC state $(\epsilon=0)$. This
is the advantageof considering the HCS over the SPAC state.

\section{{\normalsize{}Higher-order antibunching\label{sec:Higher-order-antibunching}}}

\begin{figure}
\begin{centering}
\includegraphics{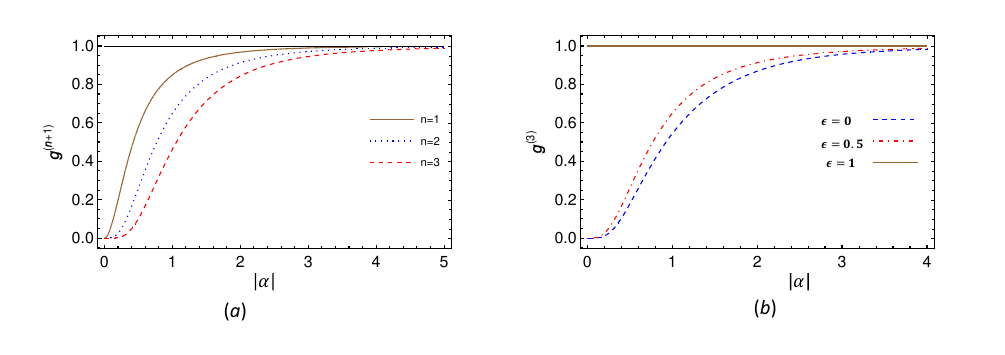}
\par\end{centering}
\caption{(Colour online) Plot of higher-order antibunching with $\left|\alpha\right|$:
(\textit{a}) for $n=1,$$2$ and $3$ with $\epsilon=0.5$; (\textit{b})
for $\epsilon=0,\,0.5,\,1,$with $n=2$. }
\label{Fig.4: plot of HOA}
\end{figure}

Photon antibunching is another well-known nonclassical effect of our
interest. Photon antibunching is a phenomenon where photons prefer
to be discrete rather than clustered. In this section, we investigate
the higher-order as well as the normal-order antibunching properties
of the HCS. Lee \cite{Lee-higherorder} first introduced specific
criteria for higher-order antibunching. Numerous, nearly identical
variations of this criterion have been offered over time. Pathak and
Garcia presented one such criterion \cite{pathak-higherorderanti}.
Higher-order antibunching in various optical systems \cite{hyper-Raman,kishore-cocoupler}
and atomic systems \cite{sandip_two mode_HOA,sandip-bec} is extensively
studied by some of the present authors using the Pathak and Garcia
criterion. But no effort has been made yet to investigate the existence
of higher-order antibunching in HCS. In order to do this, we use the
Pathak-Garcia criterion \cite{pathak-higherorderanti},which is for
$n^{th}$order antibunching as follows:
\begin{equation}
g^{(n+1)}=\frac{\left\langle a^{\dagger\,n+1}a^{n+1}\right\rangle }{\left\langle a^{\dagger}a\right\rangle ^{n+1}}<1,\label{eq:HOA}
\end{equation}
where $n=1$ and $n\geq$2 refer to usual antibunching, and higher-order
antibunching, respectively. Now, in order to investigate the higher-order
antibunching we obtain $\left\langle a^{\dagger}a\right\rangle $
for the HCS as 
\begin{equation}
\left\langle a^{\dagger}a\right\rangle =\mathscr{N}^{2}\left[1-\epsilon+\left(3-2\epsilon\right)\left|\alpha\right|^{2}+\left(1-\epsilon\right)\left|\alpha\right|^{4}+2\sqrt{\epsilon(1-\epsilon)}\left(1+\left|\alpha\right|^{2}\right)\,Re\left[\alpha e^{-i\varphi}\right]\right].\label{eq:a dagger a}
\end{equation}
Using Eqs. (\ref{eq:Average(a dagger n a m)}) and (\ref{eq:a dagger a})
we evaluate Eq. (\ref{eq:HOA}) to study the higher-order antibunching
of the HCS. To gain a sense of the higher-order antibunching in HCS,
we plot Eq. (\ref{eq:HOA}) with $\left|\alpha\right|$ in Fig. \ref{Fig.4: plot of HOA}
which shows that antibunching is observed in normal as well as higher
order. Fig. \ref{Fig.4: plot of HOA}(\textit{a})\textit{ }shows that
the degree of nonclassicality rises with order number. Fig. \ref{Fig.4: plot of HOA}(\textit{b})
shows how the antibunching property also depends on the mixing parameter
$\epsilon$ and it reduces into a coherent state distribution for
$\epsilon=1$. 

\section{{\normalsize{}Conclusions\label{sec:Conclusions}}}

Nonclassical properties and/or quantum states are needed for obtaining
any kind of quantum advantage. In fact, it's essential for achieving
quantum supremacy \cite{quantum-supremacy}. Further, an important
resource for achieving useful nonclassical features is the set of
engineered quantum states. Several applications of engineered quantum
states have been proposed in the recent past, but not all such states
can be produced using the available technologies. This makes the experimentally
realizable engineered quantum states having useful nonclassicality
extremely important. The importance of the present work is here, as
the present study shows that HCS is an experimentally realizable engineered
quantum state with higher-order nonclassical features. Although up
to the $6^{th}$order squeezing and $4^{th}$order antibunching have
been illustrated, our results hold for any order. It's interesting
to note that the HCS exhibits higher-order squeezing for low photon
counts, whereas the SPAC state does not. This illustrates why researching
the HCS is more important than the SPAC state. Additionally, the depth
of antibunching increases with increasing orders, indicating the necessity
of studying higher-order antibunching.

As the very clear presence of higher-order antibunching and higher-order
squeezing in HCS is obtained here following an analytical approach,
it's expected that the beauty of the analytic solution will encourage
others to perform similar studies with other engineered quantum states.
Further, HCS, being higher-order nonclassical and experimentally realizable,
is expected to be useful in quantum metrology and quantum communication
in the near future. With this hope, we conclude this paper.

\textbf{Acknowledgement:}\textcolor{red}{{} }B.S. expresses gratitude
to Anirban Pathak for some valuable technical conversations with him
and his assistance and interest in this work.

\end{document}